\documentclass[
    ,final            
  ]
  {aipproc}

\layoutstyle{6x9}

\begin{document}

\title{Flavor content of the nucleon in an unquenched quark model 
\footnote{Invited talk at XIII Mexican School of Particles and Fields, 
San Carlos, Sonora, Mexico, October 6-11, 2008, to be published in 
AIP Conference Proceedings}}


\author{R. Bijker}{
  address={ICN-UNAM, AP 70-543, 04510 M\'exico DF, M\'exico}
}

\author{E. Santopinto}{
  address={INFN, via Dodecaneso 33, 16164 Genova, Italy}
}

\begin{abstract}
We discuss the flavor content of the nucleon in an unquenched quark model in which 
the effects of quark-antiquark pairs ($u \bar{u}$, $d \bar{d}$ and $s \bar{s}$) are 
taken into account in an explicit form. It is shown that the inclusion of $q \bar{q}$ 
pairs leads to an excess of $\bar{d}$ over $\bar{u}$ quarks in the proton and to a 
large contribution of orbital angular momentum to the spin of the proton. 
\end{abstract}

\maketitle

\section{Introduction}

The flavor content of the nucleon provides an important test for models of nucleon 
structure. A first glimpse of the composite nature of the nucleon originated from the 
large values of the magnetic moment of the proton and neutron \cite{hess}. The ratio 
of magnetic moments $\mu_p/\mu_n$ is one of the early successes of the constituent 
quark model in which the nucleon is described in terms of three valence quarks \cite{beg}. 

Evidence for the importance of sea quarks is provided by studies of the flavor asymmetry 
of the nucleon sea via the measured violation of the Gottfried sum rule by the 
New Muon Collaboration (NMC) \cite{nmc} which was confirmed by other experimental 
collaborations \cite{DY,hermes}. All experiments show that there are more $\bar{d}$ 
quarks in the proton than $\bar{u}$ quarks. Recent results from parity-violating 
electron scattering experiments \cite{Spayde,Maas,Happex,Armstrong} show evidence 
for a nonvanishing strange quark contribution, albeit small, to the charge and 
magnetization distributions of the proton \cite{Young}. 

Other indications for the compositeness of the proton and its flavor content come 
from the European Muon Collaboration (EMC) which determined that only a very small 
fraction of the proton spin is carried by quarks \cite{emc}. The most recent value 
for the quark contribution to the spin of the proton is $33.0 \pm 3.9$ \% \cite{hermes1}. 

In this contribution, we address the importance of quark-antiquark pairs in the framework 
of an unquenched quark model of the nucleon, in which the contribution of quark-antiquark 
pairs is taken into account in an explicit manner. As applications we discuss the flavor 
asymmetry of the nucleon sea and the spin of the proton. 

\section{Unquenched quark model}

The impact of quark-antiquark pairs was studied by Geiger and Isgur \cite{mesons} 
in an unquenched quark model for mesons in which the $q \bar{q}$ pair is created with 
the $^{3}P_0$ quantum numbers of the vacuum.  
Subsequently, it was shown \cite{OZI} that a {\it miraculous}  
set of cancellations between apparently uncorrelated sets of intermediate states 
occurs in such a way that they compensate each other and do not destroy the good 
CQM results for the mesons. In particular, the OZI hierarchy is preserved and 
there is a near immunity of the long-range confining potential, since the change 
in the linear potential due to the creation of quark-antiquark pairs in the string 
can be reabsorbed into a new strength of the linear potential, {\em i.e.} in a new 
string tension. As a result, the net effect of the mass shifts due to pair creation 
is much smaller than the naive expectation of the order of the strong decay widths.  
However, it is necessary to sum over large towers of intermediate 
states to see that the spectrum of the mesons, after unquenching and renormalizing, 
is only weakly perturbed. An important conclusion is that no simple truncation of 
the set of meson loops is able to reproduce such results \cite{OZI}.

The extension to baryons requires a proper treatment of the permutation symmetry 
between identical quarks. As a first step, Geiger and Isgur investigated the 
importance of $s \bar{s}$ loops in the proton in a quark model to which the $s \bar{s}$ 
pair creation with quantum numbers of the vacuum is added as a perturbation \cite{baryons}. 
In the conclusions, the authors emphasized that, in order to address problems 
like the flavor asymmetry of the nucleon sea and the origin of the spin crisis, 
it is necessary to include $u \bar{u}$ and $d \bar{d}$ loops as well. 

In this contribution, we take up the challenge and present an unquenched quark model 
in which the effects of $q \bar{q}$ pairs can be studied for any initial baryon 
(ground state or resonance), for any flavor of the quark-antiquark pair (up, down 
and strange), and for any model of baryons and mesons, as long as their wave 
functions are expressed in the basis of the harmonic oscillator. 
These generalizations were made possible by two developments: the solution of the 
problem of the permutation symmetry between identical quarks by means of 
group-theoretical techniques, and the construction of an algorithm to generate 
a complete set of intermediate states for any model of baryons and mesons. 

\section{Flavor asymmetry}

The first clear evidence for the flavor asymmetry of the nucleon sea was provided 
by NMC at CERN \cite{nmc}. The flavor asymmetry is related to the Gottfried integral 
for the difference of the proton and neutron electromagnetic structure functions  
\begin{eqnarray}
S_G &=& \frac{1}{3} - \frac{2}{3} \int_0^1 dx \left[ \bar{d}(x) - \bar{u}(x) \right] ~.
\end{eqnarray}
Under the assumption of a flavor symmetric sea $\bar{d}(x)=\bar{u}(x)$ one obtains the 
Gottfried sum rule $S_G=1/3$. The final NMC value is $0.2281 \pm 0.0065$ at $Q^2 = 4$ 
(GeV/c)$^2$ for the Gottfried integral over the range $0.004 \leq x \leq 0.8$ \cite{nmc}, 
which implies a flavor asymmetric sea. The violation of the Gottfried sum rule was  
confirmed by Drell-Yan experiments \cite{DY} and a measurement of semi-inclusive 
deep-inelastic scattering \cite{hermes}. The experimental values are consistent with 
each other, even though the experiments were performed at very different values of $Q^2$. 
All experiments show evidence that there are more $\bar{d}$ quarks in the proton than 
there are $\bar{u}$ quarks. Theoretically, it was shown that in the cloudy bag model  
the coupling of the proton to the pion cloud provides a mechanism to produce a flavor 
asymmetry due to the dominance of $n \pi^+$ among the virtual configurations \cite{Thomas}.  

In the unquenched quark model, the flavor asymmetry can be calculated from the 
difference of the number of $\bar{d}$ and $\bar{u}$ sea quarks in the proton.  
Even in absence of explicit information on the (anti)quark distribution functions, 
the integrated value can be obtained directly. In a calculation with harmonic oscillator 
wave functions in which all parameters were taken from the literature \cite{baryons,CR}, 
we found that the effect of the quark-antiquark pairs on the Gottfried integral amounts 
to a reduction of about one third with respect to the Gottfried sum rule, corresponding 
to an excess of $\bar{d}$ over $\bar{u}$ quarks in the proton, in qualitative agreement 
with the NMC result. Due to isospin symmetry, the neutron has a similar excess of 
$\bar{u}$ over $\bar{d}$ quarks. 

\section{Proton spin} 

The unquenched quark model makes it possible to study the effect of quark-antiquark 
pairs on the fraction of the proton spin carried by quarks. Ever since the European 
Muon Collaboration at CERN showed that {\it the total quark spin constitutes a rather 
small fraction of the spin of the nucleon} \cite{emc}, there has been an enormous 
interest in the spin structure of the proton. The most recent value for the contribution 
of the quark spins is $33.0 \pm 3.9$ \% \cite{hermes1}. The total spin of the proton is 
distributed among valence and sea quarks, orbital angular momentum and gluons 
\begin{eqnarray}
\frac{1}{2} = \frac{1}{2} \left( \Delta u + \Delta d + \Delta s \right) + \Delta L 
+ \Delta G ~,
\end{eqnarray}
where $\Delta q$ denotes the fraction of the proton's spin carried by the light 
quarks and antiquarks with flavor $q=u$, $d$, $s$. $\Delta L$ and $\Delta G$ 
represent the contributions from orbital angular momentum and gluons, respectively. 
There is increasing evidence that the gluon contribution is small (either positive 
or negative) and compatible with zero \cite{gluonexp,gluonth}.

In the unquenched quark model, the inclusion of quark-antiquark pairs gives rise 
to a large contribution of the orbital angular momentum to the spin of the proton. 
More specifically, the $q \bar{q}$ pairs contribute about half of the proton spin, 
of which one quarter is due to the spin of the sea quarks and three quarters to 
orbital angular momentum. Similar conclusions regarding the importance of the 
contribution of orbital angular momentum to the proton spin were reached in 
studies with meson-cloud models \cite{cbm}.   

\section{Summary and conclusions}

In this contribution, we discussed the importance of quark-antiquark pairs 
in baryon spectroscopy in the framework of an unquenched quark model in which, 
the contributions of $u \bar{u}$, $d \bar{d}$ and $s \bar{s}$ loops are taken 
into account in an explicit way. As an illustration, the model was applied to the 
flavor asymmetry of the nucleon sea and the spin of the proton. The inclusion of 
$q \bar{q}$ pairs immediately leads to an excess of $\bar{d}$ over $\bar{u}$ quarks 
in the proton and to a relatively large contribution of orbital angular 
momentum to the proton spin. We note that all parameters in the calculation were 
taken from the literature, and that no attempt was made to optimize their values 
to improve the agreement with experimental data. 
 
In our opinion the first results for the proton spin and the flavor asymmetry are 
very promising and encouraging. We believe that the inclusion of the effects of  
quark-antiquark pairs in a general and consistent way, as suggested in this contribution, 
may provide a major improvement to the present CQM, increasing considerably its  
applicability. In future work, the model will be applied systematically to several 
problems in light baryon spectroscopy, such as the electromagnetic and strong 
couplings, the elastic and transition form factors of baryon resonances, 
their sea quark content and their flavor decomposition \cite{BS}.  

\begin{theacknowledgments}
This work was supported in part by Conacyt, Mexico. 
\end{theacknowledgments}

\end{document}